\documentclass[12pt]{article}          
\usepackage{graphicx}
\makeatletter  

\@addtoreset{equation}{section} \makeatother

\begin{document}

\renewcommand{\thefootnote}{\arabic{footnote}}

\newcommand{\hs}{\hspace*{0.5cm}}
\newcommand{\vs}{\vspace*{0.5cm}}
\newcommand{\be}{\begin{equation}}
\newcommand{\ee}{\end{equation}}
\newcommand{\bea}{\begin{eqnarray}}
\newcommand{\eea}{\end{eqnarray}}
\newcommand{\bary}{\begin{array}}
\newcommand{\eary}{\end{array}}
\newcommand{\bit}{\begin{itemize}}
\newcommand{\eit}{\end{itemize}}
\newcommand{\ben}{\begin{enumerate}}
\newcommand{\een}{\end{enumerate}}
\newcommand{\crn}{\nonumber \\}
\newcommand{\noi}{\noindent}
\newcommand{\al}{\alpha}
\newcommand{\la}{\lambda}
\newcommand{\bet}{\beta}
\newcommand{\ga}{\gamma}
\newcommand{\va}{\varphi}
\newcommand{\ra}{\rightarrow}
\newcommand{\om}{\omega}
\newcommand{\pa}{\partial}
\newcommand{\fr}{\frac}
\newcommand{\sm}{\sigma}
\newcommand{\bc}{\begin{center}}
\newcommand{\ec}{\end{center}}
\newcommand{\nn}{\nonumber}
\newcommand{\Ga}{\Gamma}
\newcommand{\de}{\delta}
\newcommand{\De}{\Delta}
\newcommand{\ep}{\epsilon}
\newcommand{\varep}{\varepsilon}
\newcommand{\vthe}{\vartheta}
\newcommand{\ka}{\kappa}
\newcommand{\La}{\Lambda}
\newcommand{\vr}{\varrho}
\newcommand{\si}{\sigma}
\newcommand{\Si}{\Sigma}
\newcommand{\ta}{\tau}
\newcommand{\up}{\upsilon}
\newcommand{\Up}{\Upsilon}
\newcommand{\kh}{\chi}
\newcommand{\ze}{\zeta}
\newcommand{\ps}{\psi}
\newcommand{\Ps}{\Psi}
\newcommand{\ph}{\phi}
\newcommand{\vph}{\varphi}
\newcommand{\Ph}{\Phi}
\newcommand{\Om}{\Omega}

\def\lappeq{\mathrel{\rlap{\raise.5ex\hbox{$<$}}
{\lower.5ex\hbox{$\sim$}}}}

\bc {\huge
 Neutral currents in\\

 reduced   minimal
 3-3-1 model}

\vspace*{0.5cm}

{\bf  V. T. N. Huyen$^a$, T. T. Lam$^b$,  H. N. Long$^a$,  V. Q. Phong$^c$} \\
\vspace*{0.5cm}

$^a${\it  Institute of  Physics, VAST, 10 Dao Tan, Ba Dinh, Hanoi,
Vietnam}

$^b${\it  Department of  Physics, Kien Giang Community College,
 Rach Gia, Vietnam}

$^c${\it  Department of Theoretical Physics, Ho Chi Minh City
University of Science, Vietnam}

\ec
\begin{abstract}

This work is devoted for gauge boson sector of the recently
proposed model based on $\mathrm{SU}(3)_C\otimes \mathrm{SU}(3)_L
\otimes \mathrm{U}(1)_X$ group with minimal content of leptons and
Higgses. The limits on the masses of the bilepton gauge bosons and
on the mixing angle among the neutral ones are deduced. Using the
Fritzsch anzats on quark mixing, we show that the third family of
quarks should  be different from the first two. We obtain a lower
bound on mass of the new heavy neutral gauge boson as 6.051 TeV.
Using data on branching decay rates of
the $Z$ boson,
 we  can fix the limit to the $Z$ and $Z^\prime$ mixing angle
  $\phi$ as $-0.001\le\phi\le 0.0003$.

\end{abstract}

\section{Introduction \label{introd}}

The experimental evidences of nonzero neutrino masses and mixing
\cite{pdg} have shown that the standard model (SM) of fundamental
particles and interactions must be extended. Among many extensions
of the SM known today, the models based on gauge
symmetry $\mathrm{SU}(3)_C\otimes \mathrm{SU}(3)_L \otimes
\mathrm{U}(1)_X$ (called 3-3-1 models) \cite{331m,331r} has
interesting features. First, $[\mathrm{SU}(3)_L]^3$ anomaly
cancelation requires that the number of $\mathrm{SU}(3)_L$ fermion
triplets must equal to that of antitriplets. If these multiplets
are respectively enlarged from those of the SM, the
fermion family number is deduced to be a multiple of the
fundamental color number, which is three, coinciding with the
observation. In addition, one family of quarks has to transform
under $\mathrm{SU}(3)_L$ differently from the other two. This can
lead to an explanation why the top quark is uncharacteristically
heavy.

One of the weaknesses of the mentioned 3-3-1 models that reduces
their predictive possibility is a plenty or complication in the
scalar sectors. The attempt on this direction to realize simpler
scalar sectors has recently been constructed 3-3-1  model with
minimal Higgs sector called the economical 3-3-1 model
\cite{ecn331c,ecn331r}. The 3-3-1 model with minimal content of
fermions and Higgs sector (called the reduced minimal (RM) 3-3-1
model) has also been constructed in   \cite{ecn331m}.

The aim of this work is presented in details the recently proposed
model with focus on gauge boson sector, and correct some misprints
in the original work  \cite{ecn331m}. The article is organized as
follows: In section \ref{review}, we review the basics of the
reduced minimal 3-3-1 model. Section \ref{Higgssec} is devoted for
the Higgs sector. In section \ref{sectgauge}, we give more details
on gauge bosons: their masses and mixing. Fermion masses and
Yukawa interactions (with some corrections) are given  in section
\ref{fermion}. The charged and neutral currents are presented in
section \ref{currentcn}, and using the obtained results  we get
the constraints on masses of the new neutral $Z'$ gauge boson in
section \ref{massdo}. Section \ref{zdecay} is devoted for the $Z$
decay, from which the limit on $Z - Z'$ mixing angle $\phi$ is
derived. In the last section, we summarize our main results.

\section{Particle content \label{review}}

The fermion content of the model under consideration is the same
as in the minimal 3-3-1 model \cite{331m,ecn331m}. The left-handed
leptons and quarks transform under the $SU(3)_L$ gauge group as
the triplets \bea
f_{aL} & = & \left(\begin{array}{c}\nu_{\ell_a} \\
\ell_a \\ \ell^c_{_a}\end{array}\right)_L \sim
\left({\bf 3}, 0\right) \qquad Q_{1L} =
\left(\begin{array}{c}u_1 \\ d_1 \\
T\end{array}\right)_L \sim \left({\bf 3}, \frac{2}{3}\right),
\cr Q_{i L} & = & \left(\begin{array}{c}
d_i\\
-u_i\\
D_{i }\end{array}\right)_L \sim \left({\bf 3^*},
-\frac{1}{3}\right), \label{fer} \eea
 where $a = 1,  2,  3$
and $i = 2, 3$. The $T$ exotic quark carries $5/3$ units of
positron's
electric charge, while $D_2$ and $D_3$ carry
$-4/3$ each one. In Eqs. (\ref{fer}) the numbers $0$, $2/3$, and
$-1/3$ are the U(1)$_X$ charges. The right-handed  quarks  are
singlets of the SU(3)$_L$  group, \bea && u^a_R \sim \left({\bf
1}, 2/3\right),\hs   d^a_R \sim \left({\bf 1}, -1/3\right), a= 1,
2, 3,
 \crn && T_{R}
\sim \left({\bf 1}, 5/3\right), \hs D_{i R} \sim \left({\bf 1},
-4/3\right). \eea

The charge operator is
defined by
\begin{equation}
\frac{Q}{e} = \frac{\la_3}{2} - \fr{\sqrt{3}}{2}\la_8
 + X, \label{op}\end{equation} where
$\la_3$ and $\la_8$ are the diagonal Gell-Mann matrices.
Note that for antitriplet, we have to replace the Gell-mann matrix by
$\bar \la= -\la^*$.

The scalar sector contains only two Higgs scalar triplets
\cite{ecn331m} \be
\rho = \left(\begin{array}{c}\rho^+ \\ \rho^0 \\
\rho^{++}\end{array}\right) \sim \left({\bf 3}, 1\right), \hs
\chi = \left(\begin{array}{c}\chi^- \\ \chi^{--} \\
\chi^0\end{array}\right) \sim \left({\bf 3}, -1\right).
\label{higg}\ee This  minimal content of Higgs sector is enough to
break the symmetry spontaneously and generate the masses of
fermions and gauge bosons in the model \cite{ecn331m}. The neutral
scalar fields develop the vacuum expectation values (VEVs)
$\langle\rho^0\rangle = \fr{v_\rho}{\sqrt{2}}$ and
$\langle\chi^0\rangle = \fr{ v_\chi}{\sqrt{2}}$, with $ v_\rho =
246$ GeV.

The pattern of symmetry breaking is \be {\rm SU(3)}_L\otimes{\rm
U(1)}_X \stackrel{\langle\chi^0\rangle} \longrightarrow {\rm
SU(2)}_L\otimes{\rm U(1)}_Y \stackrel{\langle \rho^0\rangle}
\longrightarrow {\rm U(1)}_{\rm em}\ee
 and so, we can expect
\be v_\chi \gg v_\rho. \label {limit}\ee
 Since lepton and
antilepton were put in the same triplet, therefore  in the model
under consideration, lepton number is not conserved.
 It is better to work
with a new conserved charge $\mathcal{L}$ commuting with the gauge
symmetry  \cite{tullyjoshi,clong} and related to the ordinary
lepton number by diagonal matrices
$L=\fr{4}{\sqrt{3}}T_8+\mathcal{L}$ .

Another useful conserved charge $\cal B$ is usual baryon number
\cite{clong} $B ={\cal B} I$. These numbers are given
\cite{tullyjoshi,clong} in the Table \ref{bnumber}
\begin{table}[h]
\caption{
   ${\cal B}$ and ${\cal L}$ charges for multiplets in
the RM 331 model.} \bc
\begin{tabular}{|c|c c c c c c c c c|}
  \hline
  Multiplet & $\chi$  & $\rho$ &  $Q_{1L}$ & $Q_{i L}$ &
$u_{aR}$&$d_{aR}$ &$T_{R}$ & $D_{i R}$ & $f_{aL}$   \\
\hline $\cal B$ charge &$0$ &
 $ 0  $ &  $\fr 1 3  $ & $\fr 1 3  $& $\fr 1 3  $ &
 $\fr 1 3  $ &  $\fr 1 3  $&  $\fr 1 3  $&
 $0  $\\
  \hline
   $\cal L$ charge &$\fr 4 3$   &
   $-\fr 2 3  $ & $-\fr 2 3  $& $\fr 2 3$ & 0 & $0$& $-2$&
 $2 $&$\fr 1 3$\\
  \hline
\end{tabular}
\label{bnumber} \ec
\end{table}

In Table \ref{lnumber}, we list particles with  non-zero lepton
number
\begin{table}[h]
\caption{
    Nonzero lepton number $L$ of fields in the RM 331 model .}
\begin{center}
\begin{tabular}{c|cc|ccc|cccc|}
    \hline
        Fields
&$l^c_L$&$l_L$ & $\rho^{++}_3$&$\chi^-_1$&$\chi^{--}_2$
&$D_{i L}$& $D_{i R}$&$T_{L}$&$T_{R}$\\
    \hline
        $L$ & $-1$ & $1$ & $-2$ & $2$&$2$&$2$&$2$&$-2$&$-2$ \\
    \hline
\end{tabular}
\label{lnumber}
\end{center}
\end{table}
Table \ref{lnumber} shows that the exotic quarks carry lepton
number two. Hence they are bilepton quarks.

\section{Higgs potential}\label{Higgssec}

The most general renormalizable scalar potential is given by
\cite{ecn331m} \bea V(\chi,\rho)&=&\mu^2_1\rho^\dagger\rho+
\mu^2_2\chi^\dagger\chi+\la_1(\rho^\dagger\rho)^2+\la_2(\chi^\dagger\chi)^2
\crn & &\mbox{}
+\la_3(\rho^\dagger\rho)(\chi^\dagger\chi)+\la_4(\rho^\dagger\chi)(\chi^\dagger\rho)\,,
\label{potential} \eea This potential is the simplest one since
the number of free parameters is reduced
 from, at least, thirteen to only six.

Expansion of  $\rho^0$ and $\chi^0$  around their VEVs is usually
\be \rho^0 \,,\, \chi^0 \rightarrow
\frac{1}{\sqrt{2}}(v_{\rho\,,\,\chi}+ R_{\rho\,,\,\chi}
+iI_{\rho\,,\,\chi}). \label{chankhong} \ee Substituting the
expansion in (\ref{chankhong}) to  the above potential we obtain
the following set of minimum constraint equations \cite{ecn331m}
\bea &&
 \mu^2_1+\la_1 v^2_\rho+\frac{\la_3 v^2_\chi}{2}=0,\crn
 &&
 \mu^2_2+\la_2 v^2_\chi+\frac{\la_3 v^2_\rho}{2}=0.
\eea
This potential immediately gives us two charged Goldstones bosons
$\rho^\pm$ and $\chi^\pm$  which are eaten by the gauge bosons
$W^\pm$ and $V^\pm$.

In the  doubly charged scalars, the  mass matrix in the basis
$(\chi^{++}\,,\,\rho^{++})$ is given by \be
\fr{\la_4}{ 2}\left(%
 \begin{array}{cc}
    v_\rho^2 &  v_\chi v_\rho \\
   v_\chi v_\rho &   v_\chi^2 \\
\end{array}%
\right)%
\ee This matrix has  the following squared mass eigenvalues \be
m^2_{\tilde h^{--}}=0 \hs \mbox{and}\hs
m^2_{h^{--}}=\frac{\lambda_4}{2}(v^2_\chi + v^2_\rho),
\label{chargmass} \ee where the corresponding eigenstates are \bea
&&\left(
\begin{tabular}{c}
$\tilde h^{++}$ \\
$h^{++} $
\end{tabular}
\ \right) = \left(
\begin{tabular}{cc}
$c_ \alpha$ & -$s_ \alpha$\\
$s_\alpha$ & $c_\alpha$
\end{tabular}
\ \right) \left(
\begin{tabular}{l}
$\chi^{++} $ \\
$\rho^{++} $
\end{tabular}
\right), \label{vcharged}
\eea
with
\be
c_\alpha = \frac{v_\chi}{\sqrt{v^2_\chi +
v^2_\rho}}\,,\,s_\alpha=\frac{v_\rho}{\sqrt{v^2_\chi + v^2_\rho}}.
\label{chargedeiggenvectors}
\ee

In  the neutral scalar sector,  in  the basis
$(R_\chi\,,\,R_\rho)$, the  mass matrix takes the following form
\be
\left(%
\begin{array}{cc}
  \la_2 v^2_\chi & \fr 1 2 \la_3 v_\chi v_\rho \\
  \fr 1 2 \la_3 v_\chi v_\rho & \la_1 v_\rho^2 \\
\end{array}%
\right)\ee This matrix gives us two eigenvalues \bea m^2_{h_1} &=&
\fr 1 2 v_\chi^2 \left(\la_1 t^2 + \la_2 - \sqrt{\De} \right),\crn
 m^2_{h_2} &=&
\fr 1 2 v_\chi^2 \left(\la_1 t^2 + \la_2 + \sqrt{\De} \right),
\eea where $t \equiv \fr{v_\rho}{v_\chi}$ and \be \De = (\la_1 t^2
- \la_2)^2 + \la_3^2 t^2. \ee The corresponding eigenvectors are \be \left(%
\begin{array}{c}
  h_1 \\
  h_2 \\
\end{array}%
\right) = \left(%
\begin{array}{cc}
  -s_\bet & c_\bet \\
  c_\bet & s_\bet \\
\end{array}%
\right)\left(%
\begin{array}{c}
  R_\chi \\
  R_\rho \\
\end{array}%
\right)\label{neurHiggs}\ee with \be c_\bet =
\fr{1}{\sqrt{2}}\left( 1 -\fr{\la_1 t^2 -
\la_2}{\sqrt{\De}}\right)^{\fr 1 2},\hs s_\bet =
\fr{1}{\sqrt{2}}\left( 1 + \fr{\la_1 t^2 -
\la_2}{\sqrt{\De}}\right)^{\fr 1 2}. \ee In the neutral
pseudoscalar sector, there are two Goldstones bosons $I_\rho$ and
$I_\chi$ which are eaten by the neutral gauge bosons $Z$ and
$Z^\prime$, respectively.

In the effective limit: $v_\chi \gg v_\rho$ we have \bea c_\al
&\approx & 1, \, s_\al \approx 0, \hs \sqrt{\De} \approx \la_2 -
\la_1 t^2 + \fr{\la_3^2}{2 \la_2}t^2,\crn
 c_\bet
&\approx & 1-\fr{\la_3^2}{8 \la_2^2}t^2,
\hs s_\bet \approx
\fr{\la_3 t}{2 \la_2}
 \eea This gives the
following consequences:\ben \item  The Goldstones boson  $\tilde
h^{--} \approx \chi^{--}$ and one physical doubly charged Higgs
boson is $h^{++}\approx \rho^{++}$.
\item Masses of neutral Higgs bosons
\be m^2_{h_1}=\left(\la_1
-\frac{\la^2_3}{4\la_2}\right)v^2_\rho,\hs
m^2_{h_2}=\la_2v^2_\chi
+\frac{\la^2_3}{4\la_2}v^2_\rho, \label{Higgsmass} \ee
\item The positiveness of masses yields: $\la_1 > 0, \, \la_2 > 0, \, 4 \la_1 \la_2 >
\la_3^2$.
\een

Let us resume content of the Higgs sector: the physical scalar
spectrum of the RM331 model is composed by a doubly charged scalar
$h^{++}$ and two neutral scalars $h_1$  and $h_2$. Since the
lightest neutral field, $h_1$, is basically a $SU(2)_L$ component
in the linear combination as in  Eq.~(\ref{neurHiggs}), we
identify it as the standard Higgs boson. Thus \be
\rho = \left(\begin{array}{c}G_{W^+} \\ \fr{v_\rho}{\sqrt{2}} + \fr{1}{\sqrt{2}}(h_1 + i G_Z)\\
h^{++}\end{array}\right), \hs
\chi = \left(\begin{array}{c}G_{V^-} \\ G_{U^{--}} \\
\fr{v_\chi}{\sqrt{2}} + \fr{1}{\sqrt{2}}(h_2 + i
G_{Z^\prime})\end{array}\right) \label{effhiggs}\ee Note that
$h^{--}$ carries  lepton number two. Hence, it is scalar bilepton.

\section{Gauge bosons}\label{sectgauge}

 The masses of  gauge bosons  appear
 in the Lagrangian part
\be
\mathcal{L=}\left( D_{\mu }\chi \right) ^{\dagger }\left(
D^{\mu }\chi \right) +\left( D_{\mu }\rho \right)
^{\dagger }\left( D^{\mu }\rho \right) , \label{deriv}
\ee where \be  D_\mu = \partial_\mu -ig A^a_\mu \frac{\la
^a}{2}-ig_X X \fr{\la_9}{2} B_\mu, \label{dhhp} \ee with $\la_9 =
\sqrt{\fr 2 3}\, \textrm{diag}(1,1,1)$ so that $\textrm{Tr}(\la_9
\la_9)=2$. The coupling constants of $SU(3)_L$ and $U(1)_X$
satisfy the following relation \be \fr{g_X^2}{g^2}= \fr{6
s_W^2}{1-4 s_W^2} \ee where  we have used the notations
$c_W=\cos\theta_W$, $s_W=\sin\theta_W$, $t_W=\tan \theta_W$
 with $\theta_W$ being the Weinberg mixing angle.
Substitution of the expansion in the Eq. (\ref{chankhong}) into
(\ref{deriv}) leads to the following result:
 The eigenstates of the charged gauge bosons and their
respective masses are given by \bea && W^{\pm}=\frac{A^1 \mp
iA^2}{\sqrt{2}}\hs \rightarrow \hs M_{W^{\pm }}^{2}
 =\frac{g^{2}v_{\rho }^{2}}{4},\label{klw} \\
&& V^{\pm}=\frac{A^4 \pm iA^5}{\sqrt{2}} \hs \rightarrow \hs
M_{V^{\pm }}^{2}
 =\frac{g^{2}v_{\chi }^{2}}{4},\crn
&&
 U^{\pm \pm}=\frac{A^6 \pm iA^7}{\sqrt{2}} \hs \rightarrow \hs M_{U^{\pm \pm }}^{2}
 =\frac{g^{2}\left( v_{\rho }^{2}+v_{\chi }^{2}\right) }{4}
\eea From (\ref{klw}), it follows that $v_\rho = 246$ GeV.  Note
that there is mass splitting of the charged gauge bosons \be
M^2_U-M^2_V=M^2_W \ee
 The covariant derivative of the lepton triplets is \bea
\frac{g}{2}\vec{\la}\vec{ A}_\mu= \left (
\begin{array}{lcr}
\frac{g}{2}(A^3_\mu+\frac{1}{\sqrt{3}}A^8_\mu)  &
\frac{g}{\sqrt{2}}W^+_\mu &
\frac{g}{\sqrt{2}}V^-_\mu  \\
 \frac{g}{\sqrt{2}}W^-_\mu & \frac{g}{2}(-A^3_\mu+\frac{1}{\sqrt{3}}A^8_\mu) &
\frac{g}{\sqrt{2}}U^{--}_\mu  \\
\frac{g}{\sqrt{2}}V^+_\mu & \frac{g}{\sqrt{2}}U^{++}_\mu &
-g\frac{1}{\sqrt{3}}A^8_\mu
\end{array}
\right ).  \eea while for the  anti-triplets we have \bea
\frac{g}{2}\vec{{\bar \la}} \vec{A}_\mu= \left (
\begin{array}{lcr}
-\frac{g}{2}(A^3_\mu+\frac{1}{\sqrt{3}}A^8_\mu)  &
-\frac{g}{\sqrt{2}}W^-_\mu &
-\frac{g}{\sqrt{2}}V^+_\mu  \\
-\frac{g}{\sqrt{2}}W^+_\mu &
-\frac{g}{2}(-A^3_\mu+\frac{1}{\sqrt{3}}A^8_\mu) &
-\frac{g}{\sqrt{2}}U^{++}_\mu  \\
-\frac{g}{\sqrt{2}}V^-_\mu & -\frac{g}{\sqrt{2}}U^{--}_\mu &
g\frac{1}{\sqrt{3}}A^8_\mu
\end{array}
\right ),  \eea where $\bar \la = -\la^*$.

In the neutral gauge boson sector, with  the basis ($A^3_{ \mu},
A^8_{\mu}, B_\mu$),  mass
 matrix  is given by
 \bea M^2 = \fr{g^2}{4}\left(%
\begin{array}{ccc}
  v^2_\rho & -\fr{ v^2_\rho}{\sqrt{3}} & -2 \kappa v_\rho^2 \\
  -\fr{ v^2_\rho}{\sqrt{3}} & \fr 1 3 (v^2_\rho + 4 v^2_\chi) &
  \fr{2}{\sqrt{3}}(v^2_\rho + 2 v^2_\chi) \\
  -2 \kappa v_\rho^2 & \fr{2}{\sqrt{3}}(v^2_\rho + 2 v^2_\chi) &
  4 \kappa^2 (v^2_\rho +  v^2_\chi) \\
\end{array}%
\right) \eea where $\kappa=\fr{g_X}{g}$. We can easily identify the
photon field $A_\mu$ as well as the massive bosons $Z$ and
$Z^\prime$ \cite{dng}
\bea A_\mu &=& s_W A^3_\mu + c_W(\sqrt{3} t_W
A^8_\mu +\sqrt{1-3t^2_W} B_\mu),\crn
 Z_\mu &=& c_W A^3_\mu -
s_W(\sqrt{3} t_W A^8_\mu +\sqrt{1-3t^2_W} B_\mu), \eea and \be
Z^\prime_\mu = -\sqrt{1-3t^2_W} A^8_\mu + \sqrt{3} t_W B_\mu \ee
where the mass-squared matrix for $\{Z, \:  Z^\prime\}$ is given
by \bea {\cal M}^2 = \pmatrix {M^2_Z & M^2_{ZZ'}\cr M^2_{ZZ'} &
M^2_{Z'}}
\end{eqnarray}
with
\bea
M^2_Z &=& \frac{1}{4}  \frac{g^2}{\cos^2\theta_W}  v_{\rho}^2,\crn
M^2_{Z'} &=& \frac{1}{3} \: g^2 \left[\frac{\cos^2\theta_W}{1\!-\!4
\sin^2\theta_W} \: v_{\chi}^2
              + \:  \frac{1\!-\!4 \sin^2\theta_W}{4\!\cos^2\theta_W}\:v_{\rho}^2
\right], \crn
M^2_{ZZ'} &=& \frac{1}{4\sqrt{3}} \, g^2
         \frac{\sqrt{1\!-\!4 \sin^2\theta_W}}{\cos^2\theta_W} v_{\rho}^2.
\eea
Diagonalizing the mass matrix gives the mass eigenstates $Z_1$ and $Z_2$
 which can be taken as mixtures,
 \bea
Z_1 & = & Z\cos \phi  - Z^\prime \sin \phi,
 \crn
  Z_2 & = & Z \sin \phi + Z^\prime \cos \phi.
  \eea
The mixing angle $\phi$  is given by
 \be  \tan  2\phi = \fr{M^2_{ Z} -  M^2_{ Z_ 1} }{
M^2_{ Z_2 } -  M^2_{ Z}}
\ee
where $M_{Z^1}$ and $M_{Z^2}$ are the {\it physical} mass eigenvalues
\bea
M^2_{Z_1}&=&\frac{1}{2}\left\{M_{Z'}^2+M_{Z}^2-[(M_{Z'}^2-M_{Z}^2)^2-
4(M_{ZZ'}^2)^2]^{1/2}\right\},\\
M^2_{Z_2}&=&\frac{1}{2}\left\{M_{Z'}^2+M_{Z}^2+[(M_{Z'}^2-M_{Z}^2)^2-
4(M_{ZZ'}^2)^2]^{1/2}\right\}.
\eea

From the symmetry breaking hierarchy, $v_\chi \gg  v_\rho$, we obtain the
lower mass bound of $Z_2$ \cite{dng}
\bea
M_{Z_2} &\displaystyle\mathop{>}_{\sim}& \sqrt{\frac{4}{3}}\:
           \frac{\cos^2\theta_W (M_{Z_2})}{\sqrt{1\!-\!4
\sin^2\theta_W(M_{Z_2})}}
              M_{Z_1} \crn
&\displaystyle\mathop{>}_{\sim}& 400 \; {\rm GeV} \ . \eea For
practical calculations, it is useful the following relations \bea
&&  A^3_\mu = c_W Z_\mu + s_W A_\mu, \crn && A^8_\mu = \sqrt{3}
t_W s_W Z_\mu + \sqrt{1-3t^2_W}Z^{\prime}_\mu -\sqrt{3}s_W
A_\mu,\crn
 &&B_\mu = - s_W
\sqrt{1-3t^2_W} Z_\mu + \sqrt{3}t_W Z^{\prime}_\mu + c_W
\sqrt{1-3t^2_W}A_\mu. \eea Trilinear and quartic interactions of
the gauge bosons are the same as in Ref.~\cite{331gauge} Using
data on the wrong muon decay \cite{pdg} \be Br (\mu \rightarrow e
+ \nu_e + \tilde{\nu}_\mu ) < 1.2 \, \% \hs \textrm{at} \hs 90 \%
\hs \textrm{CL} \ee we get a lower limit on singly charged
bilepton gauge boson as follows (see,
 the last reference  in \cite{331r})
\be M_V \ge 230\  {\rm GeV}
\ee
This means that the model works in quite lower energy limit available for example such as the
CERN LHC.

\section{Fermion masses}
\label{fermion}

As in the original minimal version, in this model,  the singlet
right-handed lepton does not exist. Thus, the fermion masses
 are due to effective operators.
The appropriate sources of mass for each fermion in the model are:
the  Yukawa couplings give the exotic quark  masses \cite{ecn331m}
\bea L^{exot}_{Yuk} & = &\la^T_{11}\bar Q_{1L}\chi T_{R} +
\la^D_{ij}\bar Q_{iL}\chi^* D_{jR} + H.c.\crn & = &\la^T_{11}
(\bar{u}_{1L} \chi^- + \bar{d}_{1L} \chi^{--} + \bar{T}_{ L}
\chi^0) T_{ R} \crn & & \la^D_{ij} (\bar{d}_{i L} \chi^+ -
\bar{u}_{i L} \chi^{++} + \bar{D}_{i L} \chi^{0*}) D_{j R}+ H.c.
\label{exoticquarks} \eea
 When the $\chi$ field develops its VEV, these
couplings lead to the mass matrix in the basis
$(T\,,\,D_2\,,\,D_3)$,
\be M_J= \fr{v_\chi}{\sqrt{2}}
\left(%
\begin{array}{ccc}
\la^T_{11} & 0 & 0 \\
0& \la^D_{22} & \la^D_{23}\\
0 & \la^D_{32} & \la^D_{33}\\
\end{array}%
\right) \ee which, after diagonalization, leads to mass
eigenvalues at $v_\chi $ around  few ~TeV scale \cite{ecn331m}.

For the ordinary quarks, their masses come from both
renormalizable Yukawa interactions and specific effective
dimension-five operators given by \bea
 - L^u_{Yuk} & = & \la^u_{ia}\bar Q_{iL}\rho^* u_{aR} + \frac{\la^u_{1a}}{\La}\
\varepsilon_{nmp}\left(\bar Q_{1Ln}\rho^*_m\chi^*_p\right)u_{aR} +
H.c.\crn
 & = & \la^u_{i a} (\bar{d}_{iL} \rho^- - \bar{u}_{i L} \rho^{0*} +
\bar{D}_{i L} \rho^{--}) u_{a R}\crn &+& \fr{\la^u_{1 a}
}{\La}\left[  \bar{u}_{1L}(\rho^{0*} \chi^{0*} -
\rho^{--}\chi^{++}) + \bar{d}_{1L}(\rho^{--} \chi^+ -\rho^{-}
\chi^{0*})\right. \crn &+& \left.  \bar{T}_{ L}(\rho^- \chi^{++}
- \rho^{0*} \chi^+) \right] u_{aR}+ H.c.
 \label{Yuku}
\eea In the basis $(u_1 \,,\,u_2\,,\,u_3)$,  the up-type quarks
mass matrix is given by
 \be m_u = \frac{v_\rho}{\sqrt{2}}
\left(
\begin{array}{ccc}
\la^u_{11}\fr{v_\chi}{\sqrt{2}\La}  & \la^u_{12}
\fr{v_\chi}{\sqrt{2}\La} & \la^u_{13}\fr{v_\chi}{\sqrt{2}\La}  \\
-\la^u_{21} & -\la^u_{22} &- \la^u_{23}\\
-\la^u_{31} & -\la^u_{32}&
-\la^u_{33}
\end{array}
\right) \ee For down quark sector, the relevant Yukawa
interactions are \bea - L^d_{Yuk} & = & \la^d_{1a}\bar
Q_{1L}\rho d_{aR} + \frac{\la^d_{ia}}{\La}
 \varepsilon_{nmp}\left(\bar Q_{iLn}\rho_m\chi_p\right)d_{aR} + H.c.\crn
 & = & \la^d_{1 a} (\bar{u}_{1L} \rho^+ + \bar{d}_{1L} \rho^{0} +
\bar{T}_{ L} \rho^{++}) d_{a R}\crn
&+& \fr{\la^d_{i a} }{\La}\left[  \bar{d}_{iL}(\rho^0 \chi^0 -
\rho^{++}\chi^{--}) + \bar{u}_{iL}(\rho^+ \chi^0
-\rho^{++} \chi^-)\right.
\crn
&+& \left.  \bar{D}_{i L}(\rho^+ \chi^{--} - \rho^0 \chi^-) \right] d_{aR}+ H.c.
 \label{Yukd}
 \eea
Thus, in the basis $(d_1\,,\,d_2\,,\,d_3)$, the mass matrix for
the down-type quarks  is \be m_d =\frac{v_\rho}{ \sqrt{2}} \left(
\begin{array}{ccc}
 \la^d_{11}  & \la^d_{12} & \la^d_{13}  \\
\la^d_{21}  \fr{v_\chi}{\sqrt{2}\La}& \la^d_{22}\fr{v_\chi}{\sqrt{2}\La}
 & \la^d_{23}\fr{v_\chi}{\sqrt{2}\La}\\
\la^d_{31}\fr{v_\chi}{\sqrt{2}\La} & \la^d_{32}\fr{v_\chi}{\sqrt{2}\La}
 & \la^d_{33}\fr{v_\chi}{\sqrt{2}\La} \\
\end{array}
\right) \ee

It was shown  that as the minimal version, this model is
 perturbatively
reliable at the scale around  $\La = 4 - 5 $ TeV ~\cite{dias05}.
In the model under consideration, there are 18 free Yukawa
couplings to generate masses for 6 quarks only.
 For a naive analysis
\cite{ecn331m}, we just take the diagonal case
where \bea m_u &\approx & \la^u_{11} \fr{v_\chi v_\rho}{2
\La}, \hs m_d \approx\la^d_{11} \fr{ v_\rho}{\sqrt{2}}, \hs
m_s \approx \la^d_{22} \fr{v_\chi v_\rho}{2 \La},\crn
 m_c &\approx & -\la^u_{22} \fr{ v_\rho}{\sqrt{2}}, \hs
  m_b \approx
\la^d_{33} \fr{v_\chi v_\rho}{2 \La}, \hs  m_t \approx
-\la^u_{33} \fr{ v_\rho}{\sqrt{2}}.\eea
 For sake of simplicity, assuming   $\La  = 5$ TeV,
  $v_\chi = 1$ TeV, $m_u = 2.5$ MeV, $m_d = 4.95$ MeV,
$m_s = 105$ MeV, $m_c = 1.26$ GeV, $m_b = 4.25$ GeV and $m_t =
173$ GeV, we get then \cite{ecn331m} $\la^u_{11} \approx
10^{-3}$, $\la^d_{11} \approx 2.8 \times 10^{-5}$,
$\la^d_{22} \approx 2.1 \times 10^{-2}$, $\la^u_{22}
\approx - 7.24 \times 10^{-3}$, $\la^d_{33} \approx 8.5 \times
10^{-1}$, $\la^u_{33} \approx - 1.03$.

 With a scale $\La \sim 4 -5 $ TeV,  to guarantee the proton stability, as in Ref.~\cite{truly},
  a discrete $Z_2$ symmetry over the quark fields
\[ Q_{aL}\rightarrow -Q_{aL}, \hs q_{aR}\rightarrow -q_{aR}, \]
should be imposed.

The following  effective five-dimension  operator will generate
masses for the charged leptons  \cite{ecn331m}
 \bea L^l_{Yuk} & =
&\frac{\kappa_l}{\La}\left(\overline{f^c_L}\rho^*\right)\left(\chi^\dagger
f_L \right) + H.c.\crn & =
&\frac{\kappa^l}{\La}\left[\overline{\nu^c_{L}} \rho^-  +
\overline{l^c_{L}} \rho^{0*} + \overline{l_{L}} \rho^{--}
\right]\left[\nu_{L} \chi^+  +l_{L} \chi^{++} + l^c_{L}\chi^{0*}
\right] + H.c.\crn &=
&\frac{\kappa^l}{\La}\left[\overline{\nu^c_{L}} \rho^-\nu_{L}
\chi^+ + \overline{\nu^c_{L}} \rho^-l_{L} \left(c_{\al}\tilde
h^{++}+s_{\al}h^{++}\right)
+\frac{1}{\sqrt{2}}\overline{\nu^c_{L}}
\rho^-l^c_{L}\left(v_{\chi}- iI_{\chi}\right)\right]\crn &+
&\frac{\kappa^l}{\La\sqrt{2}}\overline{\nu^c_{L}} \rho^-
l^c_{L} \left(c_{\beta} h_2- s_{\beta} h_1\right)\crn &+
&\frac{\kappa^l}{\La\sqrt{2}}\overline{l^c_{L}}\left(v_{\rho}-iI_{\rho}\right)
\left[\nu_{L} \chi^+ +l_{L}\left(c_{\al}\tilde
h^{++}+s_{\al}h^{++}\right)\right] \crn &+
&\frac{\kappa^l}{2\La}\overline{l^c_{L}}\left(v_{\rho}-iI_{\rho}\right)
l^c_L \left(v_{\chi}-iI_{\chi}\right) \crn &+
&\frac{\kappa^l}{2\La}\overline{l^c_{L}}\left(v_{\rho}-iI_{\rho}\right)
l^c_L \left(c_{\beta}h_2-s_{\beta}h_1\right) \crn &+
&\frac{\kappa^l}{\La\sqrt{2}}\overline{l^c_{L}}\left(c_{\beta}h_1+s_{\beta}h_2\right)
\left[\nu_{L} \chi^+ +l_{L}\left(c_{\al}\tilde
h^{++}+s_{\al}h^{++}\right)\right]\crn&+
&\frac{\kappa^l}{2\La}\overline{l^c_{L}}\left(c_{\beta}h_1+s_{\beta}h_2\right)
 l^c_L \left(v_{\chi}-iI_{\chi}\right) \crn &+
&\frac{\kappa^l}{2\La}\overline{l^c_{L}}\left(c_{\beta}h_1+s_{\beta}h_2\right)
 l^c_L\left(c_{\beta}h_2-s_{\beta}h_1\right)\crn &+
&\frac{\kappa^l}{\La}\overline{l_{L}}
\left(c_{\al}h^{--}-s_{ \al}\tilde
h^{--}\right)\left[\nu_{L} \chi^+ +l_{L}\left(c_{\alpha} \tilde
h^{++}+s_{\alpha}h^{++}\right)+\frac{l^c_L}{\sqrt{2}}
\left(v_{\chi}-iI_{\chi}\right)\right]\crn &+
&\frac{\kappa^l}{\sqrt{2}\La}\overline{l_{L}}
\left(c_{\al}h^{--}-s_{\al} \tilde
h^{--}\right)l^c_L\left(c_{\beta} h_2-s_{\beta}h_1\right) \crn &+
& H.c. \label{Yukl} \eea From (\ref{Yukl}), it follows  masses for
the charged leptons $m_l= \fr{v_\chi}{2\La}\kappa_l v_\rho \approx
\fr{1}{2}\kappa_l v_\rho $. Taking into account  $m_e = 0.5$ MeV,
$m_\mu = 105$ MeV, $m_\tau = 1.77$ GeV, one  gets \cite{ecn331m}
$k_e = 2 \times 10^{-5}$, $k_\mu = 4.3 \times 10^{-3}$ and $k_\tau
= 7.2 \times 10^{-2}$.

From (\ref{Yukl}), it follows  the interaction \be \frac{\kappa^l
c_\al v_\chi}{\sqrt{2}\La} h^{--}\overline{l_{L}}l^c_L \approx
\frac{\kappa^l c_\al }{\sqrt{2}} h^{--}\overline{l_{L}}l^c_L\ee
which is
 responsible  for lepton-number violating decay of
$h^{--}$ to two charged leptons. This would be a specific
character of the model.

Generation for correct neutrino mass, in this model, is still open
question \cite{ecn331m}.

\section{ Charged and neutral currents}
\label{currentcn}

The interactions among the gauge bosons and fermions are read off from
\bea
{\cal L}_F & = & \bar{R}i\ga^\mu(\partial_\mu - ig_X
B_\mu X )R\crn
           & + & \bar{L}i\ga^\mu \left(\partial_\mu-i\frac{g_X}{\sqrt{6}}
B_\mu X - ig\sum^8_{a=1} W^a_\mu . \frac{\la_a}{2}\right)L,
\label{current}
\eea
where $R$ represents any right-handed singlet and $L$ any left-handed
triplet or antitriplet.

The interactions among the charged vector fields with leptons are \cite{ecn331m}
\bea
{\cal L}^{CC}_l  &= &  \frac{g}{\sqrt{2}}(\bar{\nu}_{a L}\ga^\mu V^l_{PMNS} e_{a L}W^+_\mu +
\bar{e^c}_{a R} O^V \ga^\mu \nu_{a_L} V^+_\mu \crn
&+&\bar{e}_{a L}\ga^\mu e^c _{a R} U^{++}_\mu + \mbox{H.c.}).
\label{ccl}
\eea
with $V^l_{PMNS}=V_L^{\nu \dag}$ being the PMNS mixing matrix and $O^V = V_L^\nu$ is the matrix
diagonalizing  neutrino mass one.

For the quarks we have
\bea
{\cal L}^{CC}_q  &= & \frac{g}{\sqrt{2}}[\bar{u}_{L} V^q_{CKM}\ga^\mu d_{L} W^+_\mu +
(\bar{T}_{L}\ga^\mu (V_L^u)_{1 a} u_{aL}- \bar{d}_{lL}\ga^\mu (V_L^{d\dag})_{li } D_{i_{L}})V^+_\mu
\crn
                 &+&(\bar{u}_{l_L} (V_L^{u \dag})_{li }\ga^\mu D_{i_{L}} +\bar{T}_{L}\ga^\mu  (V_L^{d })_{1 a }
d_{a_L})U^{++}_\mu + \mbox{H.c.}],
\label{ccq}
\eea
where $i,l = 2,3$, $V^q_{CKM} = V_L^{u \dag}V_L^{d} $ is the CKM mixing matrix.
One assumes that the  exotic quarks come in a diagonal basis.

We can see that the interactions with the $V^+$ and $U^{++}$ bosons
violate the lepton number (see Eq.(\ref{ccl})) and the weak isospin
(see Eq.(\ref{ccq})).

The electromagnetic current for fermions is the usual one
\be
Q_fe\bar{f}\ga^\mu fA_\mu ,
\ee
where $f$ is any fermion with $Q_f=0, -1, 2/3, -1/3, 5/3, -4/3$
and the electromagnetic coupling constant   $e$ is identified as
follows
\be
e = g\sin\theta_W.
\label{egtheta}
\ee

The neutral current interactions can be written in the form
\bea
{\cal L}^{NC}&=&\frac{g}{2c_W}\left\{\bar{f}\ga^{\mu}
[a_{1L}(f)(1-\ga_5) + a_{1R}(f)(1+\ga_5)]f
Z^1_{\mu}\right.\crn
             &+&\left.\bar{f}\ga^{\mu}
[a_{2L}(f)(1-\ga_5) + a_{2R}(f)(1+\ga_5)]f Z^2_{\mu}\right\}.
\label{nc}
\eea
The couplings of fermions
with $Z^1$ and $Z^2$ bosons are given as follows
\bea
a_{1L,R}(f) &=&\cos\phi\ [T^3(f_{L,R})-s_W^2 Q(f)]\crn
           &-&\sin\phi \left[\frac{X(f_{L,R})}{ \sqrt{3} }\left(\fr{1-s_W^2}{\sqrt{ 1-4s_W^2}}\right)
           -\fr{\sqrt{ 1-4s_W^2}}{2\sqrt{3}}Y(f_{L,R}) \right],\crn
a_{2L,R}(f)&=& \cos\ph\left[\frac{X(f_{L,R})}{ \sqrt{3}
}\left(\fr{1-s_W^2}{\sqrt{ 1-4s_W^2}}\right)
           -\fr{\sqrt{ 1-4s_W^2}}{2\sqrt{3}}Y(f_{L,R}) \right]\crn
           &+&\sin\phi\ [T^3(f_{L,R})-s_W^2 Q(f)],
\label{vaz}
\eea
where $T^3(f)$ and $Q(f)$ are, respectively, the third component
of the weak isospin and the charge of the fermion $f$. Note that
for the exotic quarks, the weak isospin is equal to zero.
Eqs.~(\ref{vaz}) are valid for both left- and right-handed
currents. Since the value of  $X$ is different for triplets and
antitriplets, the $Z^2$ coupling to left-handed ordinary quarks is
different for  the first family,  and thus flavor changing.

We can also express the neutral current interactions of Eq.~(\ref{nc})
in terms of the vector and axial-vector couplings as follows
\bea
{\cal L}^{NC}&=&\frac{g}{2c_W}\left\{\bar{f}\ga^{\mu}
[g_{1V}(f)-g_{1A}(f)\ga_5\right] f Z^1_{\mu}\crn
             &+& \left.\bar{f}\ga^{\mu}
[g_{2V}(f)-g_{2A}(f)\ga_5]f Z^2_{\mu}\right\}.
\label{ncva}
\eea
The values of these couplings are
\bea
g_{1V}(f)&=&\cos\phi\ [T^3(f_L)-2 s_W^2 Q(f)]\crn
      &-&\sin\phi\left[
\frac{ X(f_L) }{\sqrt{3}} \left(\fr{1- s_W^2}{\sqrt{1-4s_W^2}}
\right) -\fr{\sqrt{ 1-4s_W^2}}{2\sqrt{3}}Y(f_{L}) +\frac{ Q(f_R)
}{\sqrt{3}} \left(\fr{3 s_W^2 }{\sqrt{1-4s_W^2}} \right)
 \right],\crn
g_{1A}(f)&=&\cos\phi\ T^3(f_L)\crn
      &-&\sin\phi\left[
\frac{ X(f_L) }{\sqrt{3}} \left(\fr{1 - s_W^2}{\sqrt{1-4s_W^2}}
\right) -\fr{\sqrt{ 1-4s_W^2}}{2\sqrt{3}}Y(f_{L}) -\frac{ Q(f_R)
}{\sqrt{3}} \left(\fr{3 s_W^2 }{\sqrt{1-4s_W^2}} \right)
 \right]\crn
\eea where we have used $Q(f_R)= X(f_R)$ for the singlets. The
values of $g_{1V}, g_{1A}$ and $g_{2V}, g_{2A}$ are listed in
Tables \ref{znumber}, where the first generation is assumed to
belong to the triplet. However, to get some indication as to why
the top quark is so heavy, we have to treat the third generation
differently from the first two  as in Refs~\cite{331m}
and~\cite{lng}. \vspace*{0.3cm}

\begin{table}[h]
\caption{The $Z^1 \rightarrow f\bar{f}$ couplings in the RM 331 model.}
\begin{tabular}{|c|c|c|}  \hline
f &$ g_{1V}(f)$ & $g_{1A}(f)$  \\  \hline $e, \mu, \tau$  &
$(-\frac{1}{2}+2s_W^2)\cos\phi-  \sin\phi  \fr{
\sqrt{3(1-4s^2_W)}}{2}$&$-\frac{1}{2}\cos\phi -  \sin\phi
\fr{\sqrt{(1-4s^2_W)}}{2\sqrt{3}} $\\  \hline $\nu_e, \nu_{\mu},
\nu_{\tau}$ &$\frac{1}{2}(\cos\phi-\sin\phi \fr{(1-4s^2_W)^{1/2}}{
\sqrt{3}})$ & $\frac{1}{2}(\cos\phi-\sin\phi
\fr{(1-4s^2_W)^{1/2}}{ \sqrt{3}})$
 \\ \hline t
&$(\frac{1}{2}-\frac{4s_W^2}{3})\cos\phi-\sin\phi
 \fr{1+4s^2_W}{2\sqrt{3} (1-4s^2_W)^{1/2}}$
& $\frac{1}{2}\cos\phi-\sin\phi\fr{(1-4s^2_W)^{1/2}}{2 \sqrt{3}}
$\\ \hline b  & $(-\frac{1}{2}+\frac{2s_W^2}{3})\cos\phi -\sin\phi
 \fr{1-2s^2_W}{2 \sqrt{3(1-4s^2_W)}}$ & $-\frac{1}{2}\cos\phi- \sin\phi
 \fr{1+2s^2_W}{2 \sqrt{3(1-4s^2_W)}}
$\\  \hline u,c &$ (\frac{1}{2}-\frac{4s^2_W}{3})\cos\phi +\sin\phi
\fr{1-6s^2_W}{2 \sqrt{3(1-4s^2_W)}} $ &$\frac{1}{2}\cos\phi +
\sin\phi \fr{1+2s^2_W}{2 \sqrt{3(1-4s^2_W)}} $\\ \hline
d,s & $(-\frac{1}{2}+\frac{2s^2_W}{3})\cos\phi+ \sin\phi
 \fr{1}{2 \sqrt{3(1-4s^2_W)}}
$&$-\frac{1}{2}\cos\phi
+\sin\phi \fr{1}{2 \sqrt{3}}
(1-4s^2_W)^{1/2}$\\ \hline
$T$&$-\frac{10}{3}s^2_W\cos\phi +\sin\phi
\fr{1-11s^2_W}{2 \sqrt{3(1-4s^2_W)}}
$&$ \sin\phi
\fr{1-s^2_W}{2 \sqrt{3(1-4s^2_W)}}
$\\
\hline
$D_i$&$\frac{8}{3}s^2_W\cos\phi -\sin\phi
\fr{1-9s^2_W}{2 \sqrt{3(1-4s^2_W)}}
$&$ - \sin\phi
\fr{1-s^2_W}{ \sqrt{3(1-4s^2_W)}}
$\\
\hline
\end{tabular}\label{znumber}
\end{table}

We can realize that in the limit $\phi = 0$ the couplings to $Z^1$ of the
ordinary leptons and quarks are the same as in the SM. Furthermore,
the electric charge  defined  in Eq.~(\ref{egtheta}) agrees
with the SM. Because of this, we can test the new
phenomenology beyond the SM.

In the model under consideration, the interactions with the heavy
charged vector bosons $V^+, U^{++}$ violate the lepton number and
the weak isospin. Because of the mixing, the mass eigenstate $Z^1$
now picks up flavor-changing couplings proportional to $\sin\phi$.
However, since $Z-Z'$ mixing is constrained to be very small,
evidence of FCNC's in the 3-3-1 model  can only be probed
indirectly at present via the $Z^2$ couplings.

\section{ Flavor-changing neutral currents and mass difference of
the neutral meson systems} \label{massdo}

Let us consider the effective Lagrangian \cite{lng,dumm} \bea
\mathcal{L}^{eff}_{\Delta
S=2}=\frac{G_F}{\sqrt{2}}\frac{M^4_{W}}{M^2_ZM^2_{Z'}}
\frac{4}{\sqrt{1-4s^2_W
}}(V^{D*}_{11}V^D_{12})^2\left[\bar{d}_L\ga^{\mu}s_L\right]^2.\label{eff2}
\eea From ~(\ref{eff2}) it is straightforward to get the mass
difference \cite{dumm} \be \De m_P =  \frac{8G_F
M^4_{W}}{9\sqrt{2}M^2_ZM^2_{Z'}(1-4s^2_W)}
\mbox{Re}\left[V^{*}_{L11} V_{L1j}\right]^2 f^2_P B_P m_P,
\label{masdif} \ee where  $j =2$ for the  $K_L - K_S$ and $j = 3$
for the $B^0 - \bar{B}^0$ mixing systems. The $D^0 - \bar{D}^0$
mass difference is given by the expression for the $K^0$ system
with replace of $V^D$ by $V^U$. The $Z - Z'$ mixing angle $\phi$
was bounded and to be~\cite{dng,lv}: $ |\phi| \leq 10^{-3} $ (see
also below). Hence if $M_{Z_2}$ is in order of one hundred TeV,
the $Z - Z'$ mixing has to be taken into account.

In the usual case, the $Z-Z'$ mixing is constrained to be very
small, it can be safely neglected. Therefore FCNC's occur only via
$Z_2$ couplings. For the shorthand,  hereafter we rename  $Z_1$ to be
$Z$ and  $Z_2$ to be $Z'$.

Since it is generally recognized that the most stringent limit
from $\Delta m_K$, we shall mainly discuss this quantity. We  use
the experimental values\footnote{According to the experimental value
in Ref.\cite{pdg}, $\Delta m_{D^0}=1.44^{+0.48}_{-0.5} \times
10^{10}$ $\hbar s^{-1}$=$1.44^{+0.48}_{-0.5} \times 10^{10} \times 6.582119
\times 10^{-22}$=$9.478^{+3.159}_{-3.291}\times 10^{-12}$ Mev.} ~\cite{pdg} presented in Table \ref{knumber}
\begin{table}[h]
\caption{Experimental data of $K^0$, $D^0$ and $B^0$ meson.}
\begin{tabular}{cccc}
  \hline

  & $K $  & $D$  & $B $ \\
\hline

$\Delta m [\rm MeV]$ & $( 3.483 \pm 0.006)\times 10^{- 12}$ & $9.478^{+3.159}_{-3.291}
\times 10^{-12}$& $( 3.337 \pm 0.033)\times 10^{- 10}$\\
\hline

Mass [{\rm MeV}]& $497.614\pm 0.024$ &$1864.86 \pm 0.13\ $ &$5279.58\pm 0.17$\\
\hline

$\sqrt{B_P} f_P [{\rm MeV}]$ &$135 \pm 19$ & $200  $ \cite{dumm}&$244 \pm 26 $\\
\hline
\end{tabular}\label{knumber}
\end{table}

Following the idea of Gaillard and Lee~\cite{gali}, it is
reasonable to expect that $Z'$ exchange contributes a $\Delta m$
no larger than observed values. Substituting Table \ref{knumber} into~(\ref{masdif}) we get
\begin{itemize}
\item In $K^0-\bar{K}^0$ system,
\begin{eqnarray}
M_{Z'}& >& 1.12819\times 10^6\left[\mbox{Re}(V^{D*}_{L11}
V^D_{L12})^2\right]^{1/2} \  {\rm GeV}. \label{gh1}
\end{eqnarray}
\item In $D^0-\bar{D}^0$ system,
\bea
M_{Z'}& >& 1.96139\times 10^6\left[\mbox{Re}(V^{U*}_{L11}
V^U_{L12})^2\right]^{1/2} \  {\rm GeV}. \label{gh2}
\eea
\item In $B^0-\bar{B}^0$ system,
\bea
M_{Z'}& >& 6.78557\times 10^5\left[\mbox{Re}(V^{D*}_{L11}
V^D_{L13})^2\right]^{1/2} \  {\rm GeV}. \label{gh3}
\eea
\end{itemize}

From the present experimental data we cannot get the constraint on
$V^{U,D}_{Lij}$. These matrix elemetns are only constrainted by
the Cabibbo-Kobayashi-Maskawa matrix. However, it would seem more natural, if Higgs
scalars are associated with fermion generations, to have the
choice of nondiagonal elements depends on the fields to which the
Higgs scalars couple. By this way, the simple Fritzsch~\cite{hf}
scheme gives us
\be
V^D_{ij} \approx \left( \frac{m_i}{m_j} \right)^{1/2},
\hspace*{1cm} i < j. \label{hfr}
\ee

In the model under consideration, the first quark family
transforms differently. The quark mass eigenstate
 are $U=(u,c,t)^{T}$ and $D=(d,s,b)^{T}$. In other models, the third family transforms
 differently so the value of $\Delta m_P$ will be differently and the quark mass eigenstate
 are $U=(t,u,c)^{T}$, $D=(b,d,s)^{T}$.

Combining  (\ref{gh1}),
 and (\ref{hfr})
we get  the following bounds on $M_{Z'}$: \bea M_{Z'}& \geq &
244.957\   {\rm  TeV}, \hspace*{0.3cm} {\rm if\ the\ first\ or\
the \ second\ quark\ family\ is\ different\ (\ in\ triplet)}, \crn
M_{Z'}& \geq& 6.051 \  {\rm  TeV}, \hspace*{0.3cm} {\rm if\ the\
third\ quark\ family\ is\ different} \label{thu} \eea From
(\ref{thu}) we see that  to keep relatively low bounds on
$M_{Z'}$ the third family should be the one that is different from
the other two i.e. is in triplet.

\section{Constraints on the $Z-Z'$ mixing angle and
the $Z^2$ mass} \label{zdecay}

There are many ways to get constraints on the mixing angle $\phi$
and the $Z^2$ mass. Below we present a simple one. A constraint on
the $Z-Z'$ mixing can be followed from the $Z$  data. Hence we now
calculate the $Z$ width in this model.

The decay width of the $Z$ boson is described by \cite{lan,dima,Gu}
\be \Ga(f \bar{f}) = \fr{\rho G_F M_Z^3}{6 \sqrt{2}\pi}
N_c^f \left( \beta^2|\bar{g}_A^f|^2 +
\frac{3\beta-\beta^3}{2}|\bar{g}_V^f|^2
\right)(1+n_f)R_{EW}R_{QCD},\label{brz} \ee where
$\beta=\sqrt{1-4\frac{m_f^2}{M_Z^2}}$ \cite{Gu}, $\beta$ is very
small and we present  a result which is correct up to terms of
order $\alpha\alpha_s$: \be \Ga(f \bar{f}) = \fr{\rho G_F M_Z^3}{6
\sqrt{2}\pi} N_c^f \left( |\bar{g}_A^f|^2 R_A^f + |\bar{g}_V^f|^2
R_V^f \right)(1+n_f),\label{brz} \ee where $N^f_C$ is the color
factor and other parameters are given in Ref.  \cite{Gu}
\bea
&&\rho=1+\delta\rho, \de\rho_{f\ne b}=\frac{3G_Fm_t^2}{8\sqrt{2}\pi^2},\crn
&&\de\rho_{f= b}=-\frac{G_Fm_t^2}{2\sqrt{2}\pi^2},
 n_b=10^{-2}\left(\frac{1}{5}-\frac{m_t^2}{2m_Z^2}\right), n_{f\ne b}\sim 0.
\eea
Here $R_A^f$ and $R_V^f$ are radiator factors to account for final
state QED  and QCD corrections,
  as well as effects due to nonzero fermions masses.

The non-factorial electroweak correction is  given by
 \cite{dima} $$R_V^f = 1+ \fr{3
\al(M_Z)}{4\pi}, \hs   R_A^f = 1-6 \fr{m_l^2 }{M_Z^2}+ \fr{3
\al(M_Z)}{4\pi},$$ where $\al(M_Z)$ denotes the QED coupling
constant at the scale $M_Z$. The QCD correction is given by
 \be
R_V^f(s) = R_A^f(s) = 1 + \fr{3\al_s}{4\pi} Q^2_f
+\frac{\al_s}{\pi}+ \mathcal{O}(\al^2). \ee By assuming the masses
of all the ordinary fermions except the $t$ quark to be much
lighter than the mass of the $Z$ boson and the masses of the
exotic quarks to be much heavier than the mass of the $Z$ boson,
the total width of the $Z$ boson  is given as \bea
\Ga^{RM331}_{total}[\rm GeV] &= &2.49632 + 1.6968 \sin 2\phi+
\mathcal{O}(\sin^2\phi),\label{ztot}\\
\Gamma^{RM331}_{b\bar{b}} [\rm GeV] &= &0.377046 + 0.98375\sin 2\phi +
\mathcal{O}(\sin^2\phi),\label{zbb}\\
\Gamma^{RM331}_{hadrons} [\rm GeV] &=&1.74022 + 1.5683\sin 2\phi+
\mathcal{O}(\sin^2\phi), \label{zha} \eea where we have  used
\cite{pdg}: $G_F=1.166378.10^{-5} \, \textrm{GeV}^{-2}$ ,\,
$\alpha^{-1}(M_Z)=128.87$, \, $\alpha_s=0.1184$, \,
$\bar{s}_W^2(M_Z)=0.23116$ , $\frac{M_{\tau}}{M_Z}=1.7768/91.187$
and $m_t=173.5\, \textrm{GeV}$.

Taking the experimental result \cite{pdg}:
$\Ga_{total}[\rm GeV] =2.4952\pm 0.0023$,
 we obtain the limit for the mixing angle
\begin{equation}
-0.001\le\phi\le 0.00034.\label{phit}
\end{equation}
Next, let us consider $R_b \equiv
\frac{\Ga(b\bar{b})}{\Ga(hadrons)}$. In the model under
consideration, from (\ref{zbb}) and (\ref{zha}), we obtain
\begin{equation}
R^{RM331}_b=0.21666+0.740083\tan\phi+\mathcal{O}(\tan^2\phi).
\label{rbb}
\end{equation}
According to the experimental result $R_b=0.21629\pm 0.00066$
\cite{pdg}, we also get
\begin{equation}
-0.001397\le \phi \le 0.00038.\label{phi2}
\end{equation}
Thus, both limits of the mixing angle in (\ref{phit}) and in
(\ref{phi2}) are consistent: $|\phi | \leq 10^{-3}$. This limit is
adapted to the condition in the previous section.

\section{Summary}
\label{conc}

 In this paper, we have presented  the reduced minimal  3-3-1
model (RM 331) with most economical particle content.
Some misprints in the original version of the RM 331 model were  corrected.

The limits on the masses of the bilepton gauge bosons and on the
mixing angle among the neutral ones were deduced.

  We have studied the FCNC's
 in the  RM 331 model  arisen from
the family discrimination in this model. This gives a
reason to conclude that the third family should
be treated differently from the first two.
In this sense, the $\Delta m_K$
gives us the lower bound on $M_{Z'}$ as 6.051 TeV.

From the data on branching decay rates of
the $Z$ boson,   the $Z$ and $Z^\prime$ mixing angle
  $\phi$  lies  at  $-0.001\le\phi\le 0.0003$.

Due to the simplicity of Higgs sector,  number of the model's free
parameters is strongly reduced and that increases the
predicability. However, the price of Higgs simplicity is that there are non-renormalizable
effective operators. In addition, a problem on neutrino masses is still an open question.
We hope to return to this stuff in the near future.

 \section*{Acknowledgment}
This research is funded by Vietnam  National Foundation for
Science and Technology Development (NAFOSTED)  under grant number
103.01-2011.63.
\\[0.3cm]

\end{document}